\documentclass[11pt]{article}
\usepackage{jheppub}

\usepackage{epsfig,comment}
\usepackage{graphicx}
\usepackage{url,hyperref}
\usepackage{amsmath,amssymb,bm}
\usepackage{slashed}
\usepackage{float}
\usepackage{orcidlink}
\usepackage{physics,braket}
\usepackage{dsfont}
\usetikzlibrary{quantikz2}
\usepackage{bbold}
\usepackage{tikz}
\usepackage{amsthm}
\usepackage{yfonts}
\usepackage[english]{babel}  

\usepackage{listings}
\usepackage{xcolor}

\title{Trigonometric continuous-variable gates and hybrid quantum simulations of the sine-Gordon model}

\author[1]{Tommaso Rainaldi,}
\author[2]{Victor Ale,}
\author[3]{Matt Grau,}
\author[1,4]{Dmitri Kharzeev,}
\author[5]{Enrique Rico,}
\author[1]{\\ Felix Ringer,}
\author[1]{Pubasha Shome,}
\author[2]{George Siopsis}

\affiliation[1]{Department of Physics and Astronomy, Stony Brook University, New York 11794, USA}
\affiliation[2]{Department of Physics and Astronomy, University of Tennessee, Knoxville, TN 37996-1200, USA}
\affiliation[3]{Department of Physics, Old Dominion University, Norfolk, VA 23529, USA}
\affiliation[4]{Energy and Photon Sciences Directorate, Condensed Matter and Materials Sciences Division,\\ Brookhaven National Laboratory,
Upton, NY 11973, USA}
\affiliation[5]{CERN, Theoretical Physics Department, CH-1211 Geneva, Switzerland}

\emailAdd{tommaso.rainaldi@stonybrook.edu}
\emailAdd{vale@vols.utk.edu}
\emailAdd{mgrau@odu.edu}
\emailAdd{dmitri.kharzeev@stonybrook.edu}
\emailAdd{enrique.rico.ortega@cern.ch}
\emailAdd{felix.ringer@stonybrook.edu}
\emailAdd{pubasha.shome@stonybrook.edu}
\emailAdd{siopsis@tennessee.edu}

\abstract{Hybrid qubit–qumode quantum computing platforms provide a natural setting for simulating interacting bosonic quantum field theories. However, existing continuous-variable gate constructions rely predominantly on polynomial functions of canonical quadratures. In this work, we introduce a complementary universality paradigm based on trigonometric continuous-variable gates, which enable a Fourier-like representation of bosonic operators and are particularly well suited for periodic and non-perturbative interactions. We present an ancilla-based framework for implementing trigonometric gates with arguments given by arbitrary Hermitian functions of qumode quadratures. The protocol yields unitary gates deterministically, and non-unitary gates through probabilistic post-selection. As a concrete application, we develop a hybrid qubit–qumode quantum simulation of the lattice sine-Gordon model. Using these gates, we prepare ground states via quantum imaginary-time evolution, simulate real-time dynamics, compute time-dependent vertex two-point correlation functions, and extract quantum kink profiles under topological boundary conditions. Our results demonstrate that trigonometric continuous-variable gates provide a physically natural framework for simulating interacting field theories on near-term hybrid quantum hardware, while establishing a parallel route to universality beyond polynomial gate constructions. We expect that the trigonometric gates introduced here to find broader applications, including quantum simulations of condensed matter systems, quantum chemistry, and biological models.}

\begin{document}
\maketitle
\newpage

\section{Introduction}
\label{sec:Introduction}

Hybrid quantum computing architectures that coherently combine discrete-variable (DV) qubits with continuous-variable (CV) bosonic qumodes (harmonic oscillators) have emerged as a powerful and flexible paradigm for quantum simulation and quantum information processing. Such hybrid platforms are now experimentally available across several leading hardware modalities, including trapped ions, superconducting circuits, and cavity-QED systems, where qubits are naturally coupled to harmonic oscillator degrees of freedom~\cite{Stavenger:2022wzz,Liu:2024mbr,Araz:2024dcy,6prx-zmdz,Kemper:2025ldr}.
The physical realization of qubits and qumodes varies significantly across platforms. In trapped-ion systems, qubits are encoded in internal electronic or hyperfine states, while collective motional modes provide long-lived bosonic qumodes~\cite{RevModPhys.75.281,wineland1998experimentalissuescoherentquantumstate}.
In superconducting circuits, nonlinear Josephson elements realize qubits that are coupled to microwave resonators acting as qumodes~\cite{Blais:2020wjs}.
These hybrid DV-CV systems natively support entangling qubit-qumode interactions, Gaussian CV operations, and selected non-Gaussian resources, enabling quantum circuits that directly encode both discrete and continuous degrees of freedom in hardware~\cite{Liu:2024mbr}.

Existing notions of universality in continuous-variable quantum computing \cite{Liu:2024mbr} are typically formulated in terms of polynomial functions of the canonical quadratures $\hat{x}$ and $\hat{p}$. Within this framework, arbitrary unitary transformations are approximated via Taylor-like expansions constructed from a finite set of elementary Gaussian and non-Gaussian gates \cite{Sinanan-Singh:2023xzs,11129874,hong2025oscillatorqubitgeneralizedquantumsignal}.
While formally universal, this approach is intrinsically local in phase space: representing operators with global structure or periodic dependence generally requires high-order polynomials and correspondingly deep circuits.

In this work, we introduce and develop an alternative, and complementary, universality paradigm for bosonic quantum gates based on trigonometric functions of continuous-variable operators. Gates of the form $e^{-i t \cos \hat A}$ and $e^{-i t \sin \hat A}$, where $\hat A$ is a Hermitian operator acting on one or more qumodes, provide access to a Fourier-like operator basis rather than a polynomial one. In close analogy with the distinction between Taylor and Fourier representations in classical analysis, trigonometric gates are naturally suited to capturing periodic structure with fewer resources. We show that, when combined with standard hybrid qubit-qumode primitives, these gates enable the systematic construction of a broad class of bosonic unitaries and thus constitute a parallel route to universality for continuous-variable quantum computation.

A central technical contribution of this paper is a deterministic and unitary method for implementing trigonometric continuous-variable gates using ancillary qubits. Rather than exponentiating trigonometric operators directly, we embed the corresponding unitary operators into an enlarged qubit-qumode Hilbert space, where they are promoted to effective operators that are both Hermitian and unitary. This construction allows them to be exponentiated using ancilla-based techniques originally developed for Pauli strings. The method applies to trigonometric functions whose arguments are arbitrary Hermitian functions of the quadratures $\hat{x}$ and $\hat{p}$, and it naturally extends to the implementation of non-unitary trigonometric operators relevant for imaginary-time evolution.

From a hardware perspective, these gates rely only on hybrid qubit–qumode interactions and ancilla-qubit control gates that are already available in hybrid architectures. These operations include conditional displacements, single-qubit rotations, and qubit-mediated phase interactions, which have been demonstrated in trapped-ion and superconducting qubit platforms. The main practical considerations are finite coupling strengths, qubit and qumode coherence times, and effective truncation of the bosonic Hilbert space due to anharmonicity and finite squeezing. Since the trigonometric gates are constructed from standard hybrid operations, they are compatible with existing experimental devices.

Beyond their algorithmic interest, trigonometric continuous-variable gates are directly motivated by quantum field theory, where cosine interactions arise ubiquitously. A paradigmatic example is the sine-Gordon model, whose Hamiltonian contains a non-polynomial cosine potential and which plays a central role in condensed-matter physics, high-energy physics, and statistical mechanics. See Ref.~\cite{Roy:2020ppa,Horvath:2021vlx,Wybo:2022chq,Kormos:2022rtu,Wybo:2023xal} for qubit-based and analog quantum simulation protocols. In $1+1$ dimensions, the sine-Gordon model is integrable and admits exact solutions with solitonic excitations in the form of kinks and anti-kinks. Through bosonization, it is closely related to fermionic theories such as the massive Thirring model \cite{Coleman:1974bu}, and it also emerges in the study of large-$N$ $\mathrm{QCD}_2$ \cite{Witten:1979kh}, where baryons can be described as solitonic configurations governed by an effective sine-Gordon dynamics, see \cite{Florio:2022uvd} and references therein. This integrability is a special feature of the 1+1-dimensional sine-Gordon model and is lost in higher dimensions: its straightforward extensions to 2+1 and 3+1 dimensions define non-integrable interacting field theories \cite{Manton:2004tk}. In these settings, the topological excitations acquire a different geometric character, with the particle-like kinks of the two-dimensional theory generalizing to line-like defects in 2+1 dimensions, which play an important role in a variety of condensed-matter systems \cite{POLYAKOV1977429, Sachdev_2011}. In 3+1 dimensions, these configurations further extend to domain-wall solutions, which have been studied extensively in high-energy and cosmological contexts as effective descriptions of topological defect structures, including models related to cosmic strings and axion dynamics \cite{Vilenkin:2000jqa,Preskill:1982cy}.

In this work, we demonstrate how the lattice sine-Gordon Hamiltonian can be mapped onto a hybrid qubit-qumode architecture, with one bosonic mode per lattice site and qubit ancillas mediating the required trigonometric interactions. Using this framework, we investigate several observables that probe both equilibrium and dynamical properties of the model: 
(i) preparation of the ground state via quantum imaginary-time evolution using non-unitary trigonometric gates;
(ii) real-time dynamics under unitary evolution;
(iii) time-dependent vertex two-point correlation functions encoding non-perturbative information about solitonic excitations; and 
(iv) quantum kink profiles obtained by imposing topological boundary conditions and analyzing the resulting ground-state expectation values and fluctuations.

We note that cosine interactions also arise naturally in quantum simulations of lattice gauge theories, where they act on gauge-field degrees of freedom. In those settings, however, the arguments of the trigonometric functions correspond to compact continuous variables associated with the underlying gauge group, whereas in the present work, we focus on non-compact continuous quadratures. In Refs.~\cite{Jha:2023ecu,Ale:2024uxf,Ale:2025sxz,Crane:2024tlj}, several strategies were explored to effectively realize compact continuous variables for lattice gauge theories, including the use of squeezing operations and the introduction of penalty terms in the Hamiltonian.

Taken together, our results demonstrate that trigonometric continuous-variable gates provide a natural and efficient language for simulating interacting bosonic quantum field theories on near-term hybrid quantum hardware, while establishing a broader gate-level framework that complements existing polynomial-based approaches to continuous-variable universality.

Throughout this work, we use the term universality in the standard continuous-variable sense of dense approximability of target unitaries on bounded regions of phase space, rather than in a strict fault-tolerant or complexity-theoretic sense. The trigonometric gate constructions introduced here do not replace the established polynomial (Taylor-based) universal gate sets, but rather complement them by providing access to a Fourier-type operator basis that is better matched to periodic and non-local structures. In this sense, trigonometric continuous-variable gates define a parallel universality paradigm: they expand the expressive operator space accessible to hybrid qubit–qumode circuits while remaining constructible using standard universal hybrid primitives. The focus of this work is on expressivity and physical suitability for non-polynomial interactions, rather than on optimal gate counts or asymptotic complexity guarantees.

The remainder of this work is organized as follows. In section~\ref{sec:2}, we introduce the notation used throughout this work and the elementary qubit-qumode gates. We also discuss notions of universality, focusing on how continuous-variable operators can be approximated using either polynomial Taylor expansions or trigonometric Fourier-type expansions. In section~\ref{sec:exponentiation}, we describe the exponentiation of Hermitian qubit-qumode operators that enables the construction of trigonometric gates. 
As a concrete application, section~\ref{sec:sineGordon} presents a quantum simulation protocol for the lattice sine-Gordon model. Using classical simulations, we explore both equilibrium and dynamical properties of the model. Finally, we summarize our findings and provide an outlook in section~\ref{sec:conclusions}.

\section{Hybrid continuous-discrete variable quantum computing}\label{sec:2}

\subsection{Notation and elementary gates}

Throughout this work, the canonical quadrature operators of a qumode are denoted by $\hat x$ and $\hat p$, satisfying
\begin{equation}
[\hat x,\hat p]= i \,,
\end{equation}
with $\hbar=1$. A single or multiple qumodes span the CV Hilbert space $\mathcal H_{\mathrm{CV}}$, which can be represented by the eigenstates $\ket{x}$ or $\ket{p}$ of the corresponding quadratures. Alternatively, the Fock basis can be used to represent qumodes with states $\ket{n}=(\hat a^\dagger)^n\ket{0}/\sqrt{n!}$, where $\ket{0}$ is the qumode ground state in the Fock basis and $\hat a=(\hat x+i\hat p)/\sqrt{2}$ denotes the bosonic annihilation operator with $[\hat a,\hat a^\dagger]=1$. Qubits can be described by the computational basis states $\ket{0}$ and $\ket{1}$ and the associated Pauli operators $\sigma\in\{X,Y,Z\}$. We also introduce the ladder operators
\begin{equation}
X^\pm=\frac{1}{2}(X\pm iY)\,,
\end{equation}
which often appear in qubit-qumode hybrid gates.

A hybrid system consisting of $N_d$ qubits and $N_c$ qumodes is described by the tensor-product Hilbert space
\begin{equation}
\mathcal H_{\mathrm{hybrid}}=\mathcal H_{\mathrm{DV}}\otimes\mathcal H_{\mathrm{CV}} \,.
\end{equation}
Operators acting on this hybrid space can couple the discrete and continuous sectors. A general hybrid operator can be written in the form
\begin{equation}\label{eq:PauliPolynomial}
\hat O = P \otimes f(\hat x,\hat p)\,,
\end{equation}
where $P$ denotes a Pauli string acting on the qubit register and $f(\hat x,\hat p)$ is a polynomial of the quadrature operators. 

Hybrid qubit-qumode platforms admit a universal description in terms of a finite set of elementary gates acting on the combined discrete and continuous degrees of freedom. A convenient choice of a universal hybrid gate set is given by~\cite{Lloyd:1998jk,Liu:2024mbr}
\begin{equation}\label{eq:universal}
\mathcal G_{\mathrm{hybrid}}
=
\Bigl\{
R_{x,y,z},
\mathrm{CNOT},
D(\xi),
S(z),
V(\gamma),
\mathrm{BS}(z),
\mathrm{CD}(\alpha)
\Bigr\}\,,
\end{equation}
where $R_i$ denote single-qubit rotations and $\mathrm{CNOT}$ is the controlled-NOT gate acting on the qubit sector. The remaining elements of $\mathcal G_{\mathrm{hybrid}}$ act on the continuous-variable sector or couple qubits and qumodes. Explicitly, these gates are defined as
\begin{align}
D(\xi) &= \exp\!\left(\xi \hat a^\dagger - \xi^* \hat a \right), \\
S(z) &= \exp\!\left[\frac{1}{2}\left(z^* \hat a \hat a - z \hat a^\dagger \hat a^\dagger \right)\right], \\
\mathrm{BS}(z) &= \exp\!\left(z \hat a^\dagger \hat b - z^* \hat a \hat b^\dagger \right), \\
V(\gamma) &= \exp\!\left(i\frac{\gamma}{3}\hat q^3\right), \\
\mathrm{CD}(\alpha) &= \exp\!\left[(\alpha \hat a^\dagger - \alpha^* \hat a)\, Z \right],
\end{align}
where $\hat a,\hat b$ denote bosonic annihilation operators corresponding to different qumodes. The gate set in Eq.~(\ref{eq:universal}) enables the construction of arbitrary unitary operators $\hat O$ of the form given in Eq.~(\ref{eq:PauliPolynomial}) through composition and commutation. While $\mathcal G_{\mathrm{hybrid}}$ is universal in this sense, it is neither minimal nor unique. Depending on the hardware platform, different universal gate sets may be advantageous.

As a concrete example, trapped-ion systems can host qubits in long-lived electronic or hyperfine states, and continuous-variable degrees of freedom in the collective motional modes of vibration of a multi-ion crystal in the trap. In this setting, qubit–qumode hybrid interactions are naturally realized through state-dependent forces generated by laser-driven electronic transitions that are detuned near a motional mode of the trap. These state-dependent forces produce qubit-conditioned displacements of the motional modes in phase space, allowing controlled trajectories whose geometry can be tailored through the amplitude, phase, and duration of the applied laser fields. By designing closed phase-space loops, the motional degrees of freedom can be disentangled from the qubit at the end of the interaction while the qubit–qumode system acquires a well-defined geometric phase, providing a physical mechanism for conditional phase operations and ancilla-mediated bosonic gates.

\subsection{Universality}

In CV quantum computations, the standard gate set naturally generates polynomials in the quadratures $\hat{x}$ and $\hat{p}$, and arbitrary operators are typically approximated via Taylor-like polynomial expansions leading to the notion of universality discussed in Refs.~\cite{Lloyd:1998jk,Liu:2024mbr}. While universal, this approach is inherently local: high accuracy over a finite phase-space region often requires high-degree polynomials, which in turn requires an increasingly larger circuit depth. The introduction of trigonometric gates such as $e^{-it\cos(c \hat{x})}$ or $e^{-it \sin(c \hat{x})}$, where $c$ is a constant, allows for a Fourier-type representation of operators. These gates are suitable to efficiently approximate target unitaries that are periodic functions or those with bounded support in phase space, with faster (potentially exponential) convergence in the number of terms. In this sense, polynomial CV gates are akin to a Taylor expansion, while trigonometric gates provide access to a Fourier basis that is better matched to global or periodic features of the desired unitary. We emphasize that the polynomial and trigonometric representations are not mutually exclusive, and that the latter relies on the former for its constructive implementation.

Using the hybrid circuits described below, one can generate operators corresponding to arbitrary trigonometric functions whose arguments are polynomials of the quadrature operators. Consider a generic analytic function of the quadratures
\begin{equation}
    f(\hat{x},\hat{p}) = \sum_{n,m=0}^{\infty}c_{nm}\hat{x}^n\hat{p}^m.
\end{equation}
The trigonometric gates introduced in this paper allow, in principle, for the construction of any unitary, whose Hermitian exponent is
\begin{align}
    g\left(\hat{x},\hat{p}\right)\equiv &\, 
    \sum_{n=0}^{\infty}\gamma_{n,0}\cos^n(f_c(\hat{x},\hat{p})) + \sum_{n=0}^{\infty}\gamma_{0,n}\sin^n(f_s(\hat{x},\hat{p})) \nonumber \\
    &+ \sum_{n,m=1}^{\infty}\gamma_{nm}\left[\cos^n(f_{c}(\hat{x},\hat{p}))\sin^m(f_{s}(\hat{x},\hat{p})) + \sin^m(f_{s}(\hat{x},\hat{p}))\cos^n(f_{c}(\hat{x},\hat{p}))\right],
\end{align}
provided that one can first construct qubit-controlled unitaries of $e^{i f(\hat{x},\hat{p})}$, where $f_c$ and $f_s$ denote two different functions. Here, $\gamma_{nm}\in \mathbb{R}$. It is important to note that the Fourier-based universality does not replace the earlier (Taylor-like) universality statement. In fact, the latter is leveraged to build the desired Fourier decomposition. A concrete example will be given in section~\ref{sec:cosgate}, where we use the conditional displacement, whose exponent is linear in $\hat{x}$, to produce the trigonometric gate $e^{-i t\cos(c \hat{x})}$.

\section{Exponentiation of qubit-qumode Hermitian operators}
\label{sec:exponentiation}

We start by reviewing a method that allows for the construction of a unitary operator where any Pauli string is exponentiated with the help of an ancillary qubit. Subsequently, we will extend this method to hybrid qubit-qumode operations.

\subsection{Exponentiation of operators using ancilla qubits}

\begin{figure}[t]
\centering
\begin{quantikz}
\lstick{$\ket{0}$} & \gate{H} & \ctrl{1} & \gate{R_x(t )}  & \ctrl{1} & \gate{H} &  \\
\lstick{$\ket{\psi}$} & \qw     & \gate{P} & \qw     & \gate{P^\dagger} & \qw     & \qw
\end{quantikz}
\caption{Circuit that implements the exponentiation of any Pauli string $P = P^\dagger$.}
\label{fig:exp_Pauli}
\end{figure}
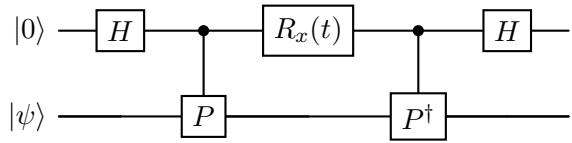

A Pauli string $P\in \{\mathbb{1}, X, Y, Z\}^{\otimes n}$ is a tensor product of Pauli operators that are both Hermitian and unitary. Because of this feature, they can be exponentiated using a single ancilla qubit. See Ref.~\cite{PhysRevA.65.032312,PhysRevLett.118.070501,Echevarria:2020wct}. The circuit in Fig.~\ref{fig:exp_Pauli} is both deterministic and unitary, which implements the unitary where the Pauli string is exponentiated
\begin{equation}
    e^{-i\frac{t}{2}P\otimes Z_a}\ket{\psi}\ket{0}_a = e^{-i\frac{t}{2}P}\ket{\psi}\ket{0}_a\,,
    \label{eq:exp_Pauli_unitary}
\end{equation}
where $\ket{0}_a$ is the ancilla qubit. Here, the Pauli string acts on a register of qubits denoted by $\ket{\psi}$, and $Z$ acts on the ancillary qubit. The result is deterministic because the ancilla qubit decouples from the rest of the circuit and does not need to be measured. In addition, it is unitary because, crucially, any Pauli string is both unitary and Hermitian. 

A naive approach to implementing trigonometric gates would be to directly exponentiate an operator of the form
$U = e^{i \hat A}$, where $\hat A$ is a continuous-variable operator. However, this strategy fails because $U$ is generally non-Hermitian, and therefore cannot serve as a valid generator for unitary time evolution. Instead, we embed the operator $U$ into a larger Hilbert space by entangling it with an ancilla qubit. This construction promotes the effective generator to a Hermitian operator acting on the joint qubit-qumode system, ensuring that the resulting evolution is both unitary and physically implementable. In this way, the procedure naturally generalizes the exponentiation of Pauli strings familiar from qubit-based quantum simulation to the case of continuous-variable operators. We introduce the following two operators
\begin{align}
    \Sigma &\, = e^{i \hat{A}\otimes X}\cdot (\mathbb{1}\otimes Z) = \frac{1}{2}\begin{pmatrix}
        U+U^\dagger & U^\dagger-U\\
        U-U^\dagger & -U - U^\dagger
    \end{pmatrix} \ , \\
    \overline{\Sigma} &\, = (\mathbb{1}\otimes Z) \cdot e^{i\hat{A}\otimes X} = e^{-i\hat{A}\otimes X}\cdot (\mathbb{1}\otimes Z) = \frac{1}{2}\begin{pmatrix}
        U+U^\dagger & U-U^\dagger\\
        U^\dagger-U & -U - U^\dagger
    \end{pmatrix}.
    \label{eq:Sigma_definition}
\end{align}
The choice of the above construction is not unique. In fact, equivalent Hermitian and unitary hybrid operators can be obtained by replacing $X \mapsto P_1$ and $Z \mapsto P_2$ with two distinct Pauli operators $P_1 \neq P_2$. For brevity, we do not list all such possibilities and instead restrict ourselves to the specific choices given in Eq.~\eqref{eq:Sigma_definition}, which are sufficient for the discussion that follows.
\begin{figure}[t]
        \centering
        \begin{quantikz}
\lstick{$\ket{0}$} & \gate{H} & \ctrl{1} & \gate{R_x(t )}  & \ctrl{1} & \gate{H} &  \\
\lstick{$\ket{\psi}$} & \qw     & \gate[wires=2]{\,\Sigma\,} & \qw     & \gate[wires=2]{\Sigma^\dagger} & \qw     & \qw\\
\lstick{$ \ket{\phi}$} \setwiretype{b}& \qw     &  & \qw     &  & \qw     & \qw
\end{quantikz}
        \caption{Circuit that implements the exponentiation of the hybrid qubit-qumode operator $\Sigma = \Sigma^\dagger$. Here, the qubit (qumode) is represented by a single (thick or triple) wire. }
        \label{fig:exp_Sigma}
\end{figure}
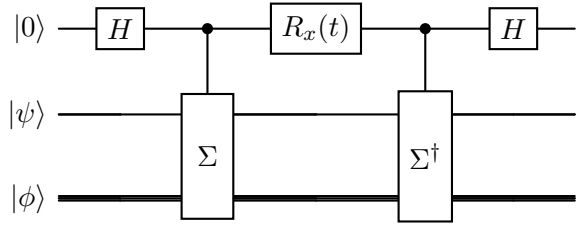
We can directly verify that $\Sigma = \Sigma^\dagger = \Sigma^{-1}$, and similarly for $\overline{\Sigma}$. We can therefore safely use the circuit shown in Fig.~\ref{fig:exp_Pauli} to exponentiate $\Sigma$ (or $\overline{\Sigma}$) by making the replacement $P \mapsto \Sigma$ (or $\overline{\Sigma}$). The resulting circuit remains both deterministic and unitary.  Note that two ancillary qubits are required: the first, labeled $a$, entangles the unitary $U$ in order to construct $\Sigma$, while the second, labeled $b$, implements the standard exponentiation procedure that places $\Sigma$ in the exponent. Analogously to the circuit in Fig.~\ref{fig:exp_Pauli}, which requires a controlled-$P$ gate, the present algorithm requires a controlled-$\Sigma$ gate, denoted by $\mathrm{C}\Sigma$. For the specific choice of $\Sigma$ given in Eq.~(\ref{eq:Sigma_definition}), we find
\begin{equation}
    \begin{split}
        \text{C}\Sigma\equiv e^{i\hat{A}\otimes X_a\otimes \Pi_{-}^b}\left(\mathbb{1}\otimes \text{C}Z_{ab}\right),\quad \text{C}\overline{\Sigma}\equiv e^{-i\hat{A}\otimes X_a\otimes \Pi_{-}^b}\left(\mathbb{1}\otimes \text{C}Z_{ab}\right),\quad \Pi_-\equiv \frac{\mathbb{1}-Z}{2}\,,
    \end{split}
\end{equation}
with
\begin{equation}
    \text{C}Z_{ab} = e^{i\pi \Pi_{-}^a\otimes \Pi_{-}^b} = H_b\,\text{CNOT}_{a\to b}\,H_b\,.
\end{equation}
Note that, up to this point, the construction is exact and involves no approximations or Trotterization errors. The corresponding circuit is shown in Fig.~\ref{fig:exp_Sigma}. There, $\ket{\psi}$ and $\ket{\phi}$ denote the initial states of the second ancillary qubit (labeled by $a$) and the bosonic mode, respectively.

\subsection{Trigonometric continuous-variable gates}

To build trigonometric continuous-variable gates whose argument can be any Hermitian operator, we first note that
\begin{equation}
    \begin{split}
        \Sigma + \overline{\Sigma} &= 2\cos\hat{A}\otimes Z,\\
        \Sigma - \overline{\Sigma} &= 2\sin\hat{A}\otimes Y.
    \end{split}
\end{equation}
Therefore, if we concatenate the circuit in Fig.~\ref{fig:exp_Sigma} for both $\Sigma$ and $\overline{\Sigma}$, we can obtain the combinations
\begin{equation}
    \begin{split}
        \mathcal{C}_{\Sigma\overline{\Sigma}}&\equiv e^{i\frac{t}{2}\Sigma}\cdot e^{i\frac{t}{2}\overline{\Sigma}} = e^{i\frac{t}{2}\left(\Sigma + \overline{\Sigma}\right)} + \mathcal{O}(t^2) = e^{it\cos\hat{A}\otimes Z_a}+ \mathcal{O}(t^2),\\
        \mathcal{S}_{\Sigma\overline{\Sigma}}&\equiv e^{i\frac{t}{2}\Sigma}\cdot e^{-i\frac{t}{2}\overline{\Sigma}} = e^{i\frac{t}{2}\left(\Sigma - \overline{\Sigma}\right)} + \mathcal{O}(t^2) = e^{it\sin\hat{A}\otimes Y_a}+ \mathcal{O}(t^2).
    \end{split}
    \label{eq:trigGates_Trotte1}
\end{equation}
By initializing the ancilla qubit in the $+1$ eigenstate $\ket{0}_a$  of $Z$ for the first expression and the $+1$ eigenstate $S_aH_a\ket{0}_a$ of $Y$ for the second, we achieve the implementation of trigonometric gates both unitarily and deterministically provided that $t \ll 1$. We can always improve the protocol by removing higher-order corrections using the higher-order Trotter-Suzuki formula \cite{SUZUKI1990319,Wiebe:2008cbb}. 

Our construction is valid for any unitary $U$, or equivalently, any Hermitian operator $\hat{A}$ for which the controlled version exists or can be constructed. Having the control-$U$ gate $e^{i\hat{A}\otimes P}$, with $P$ any Pauli operator, allows for the implementation of trigonometric gates as a function of $\hat{A}$. Of particular interest for applications are bosonic gates, where $\hat{A}$ is a function of the quadratures $\hat{x}$ and $\hat{p}$ as mentioned in section~\ref{sec:2}. 

\subsection{Example: Cosine gate of the position quadrature operator}\label{sec:cosgate}

\begin{figure}[t]
    \centering
    \begin{quantikz}[row sep=0.5cm, column sep=0.4cm]
\setwiretype{b}\lstick{$\ket{\phi}$} & \qw       & \gate[wires=2]{\text{CD}} & \qw       & \qw       & \qw       & \qw       &      \gate[wires=2]{\left(\text{CD}^\dagger\right)^2}        & \qw       &  & &  &\gate[wires=2]{\text{CD}}& &\\
\lstick{$\ket{0}_a$} & \gate{R_y^\dagger} & \qw       & \targ{}         & \qw &      & \targ{}  & \qw       &       \targ{} &  & & \targ{} & &\gate{R_y} &\\
\lstick{$\ket{0}_b$} & \gate{H}  & \qw       &\ctrl{-1}   &  & \gate{R_x(t)} & \ctrl{-1}    &          & \ctrl{-1}   &  & \gate{R_x(t)} & \ctrl{-1} & & \gate{H} &
\end{quantikz}
    \caption{Circuit implementing the cosine gate as a function of the position quadrature $\hat{x}$ up to errors of order $\mathcal{O}(t^2)$, see Eq.~(\ref{eq:cosinegate}).}
    \label{fig:cosineGate_circ}
\end{figure}
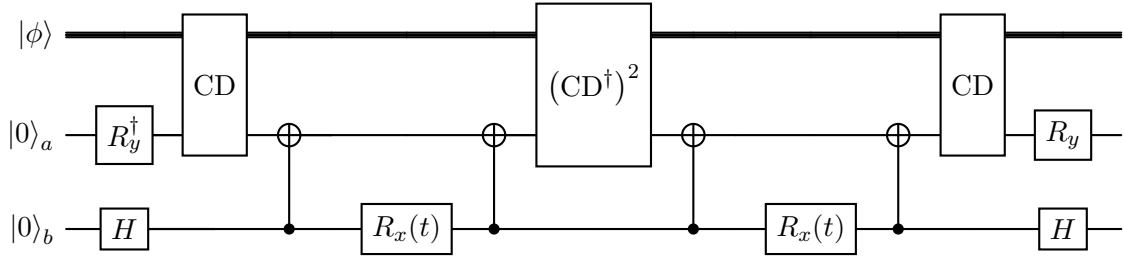

For concreteness, we construct the exact circuit decomposition for the continuous-variable cosine gate as a function of the position quadrature operator
\begin{equation}\label{eq:cosinegate}
    e^{-i t \cos(c\hat{x})}\,,
\end{equation}
where $c\in\mathbb{R}$ is a constant. We use Eq.~\eqref{eq:trigGates_Trotte1} with $\hat{A} = c\hat{x}$. We follow the construction
\begin{equation}
   H_b\,\text{C}\Sigma\, e^{-i\frac{t}{2} X_b}\,\text{C}\Sigma \text{C}\overline{\Sigma}\, e^{-i\frac{t}{2} X_b}\,\text{C}\overline{\Sigma}\,H_b\ket{0}_b\ket{0}_a\ket{\phi} = e^{-it\cos(c\hat{x})}\ket{0}_b\ket{0}_a\ket{\phi} + \mathcal{O}(t^2)\,,
\end{equation}
where we include two ancilla qubits and the qumode state that the cosine gate acts on is denoted by $\ket{\phi}$. To build $\text{C}\Sigma$ we start with a controlled displacement that we then rotate so that the control is $X_a$, i.e.
\begin{equation}
    \text{CD}_{a}\left(\frac{ic}{2\sqrt{2}}\right) = e^{ic\frac{\hat{x}}{2}\otimes Z_a\otimes \mathbb{1}_b}\overset{Z_a\to X_a}{\mapsto} R_{y_a}\text{CD}_{a}\left(\frac{ic}{2\sqrt{2}}\right)R_{y_a}^\dagger = e^{ic\frac{\hat{x}}{2}\otimes X_a\otimes \mathbb{1}_b}.
    \label{eq:CD_forCosgate}
\end{equation}
The above is a single-qubit-qumode operation but we have added the identity on qubit $b$ for later convenience.
We also need a double-controlled displacement, which can be implemented as
\begin{equation}
    \text{CCD}^\dagger_{ab}\left(\frac{ic}{2\sqrt{2}}\right) = e^{-ic\frac{\hat{x}}{2}\otimes Z_a\otimes Z_b}\overset{Z_a\to X_a}{\mapsto} R_{y_a}\text{CCD}^\dagger_{ab}\left(\frac{ic}{2\sqrt{2}}\right)R_{y_a}^\dagger = e^{-ic\frac{\hat{x}}{2}\otimes X_a\otimes Z_b},
    \label{eq:CCD_forCosgate}
\end{equation}
where we have conveniently defined
\begin{equation}
    R_{y_a}\equiv R_{y_a}(\pi/2) = e^{-i\frac{\pi}{4}Y_a}.
\end{equation}
To realize this gate, we use 
\begin{equation}
   \text{CCD}_{ab}\left(i\frac{c}{\sqrt{2}}\right)= e^{is\hat{x}\otimes Z_a\otimes Z_b} = \left(\mathbb{1}\otimes\text{CNOT}_{b\to a}\right)\text{CD}_{a}\left(i\frac{c}{\sqrt{2}}\right)\left(\mathbb{1}\otimes\text{CNOT}_{b\to a}\right),
\end{equation}
where we have used the identity
\begin{equation}
    Z_a\otimes Z_b = \text{CNOT}_{b\to a}(Z_a\otimes \mathbb{1}_b) \text{CNOT}_{b\to a}.
\end{equation}
Finally, we obtain
\begin{equation}
    \text{C}\Sigma = R_{y_a}\text{CD}^\dagger_{a}\text{CCD}_{ab}R^\dagger_{y_a}H_b\,\text{CNOT}_{a\to b}\,H_b.
\end{equation}
The construction of $\text{C}\overline{\Sigma}$ follows a similar logic.
After operator simplifications, the qubit-qumode hybrid circuit that implements the desired cosine gate in the position quadrature operator (up to higher-order corrections) is given in Fig.~\ref{fig:cosineGate_circ}, where we have neglected the argument of the controlled displacements in Eqs.~\eqref{eq:CD_forCosgate}-\eqref{eq:CCD_forCosgate} for brevity.

\subsection{Extension to non-unitary trigonometric gates}\label{sec:nonU_trigs}

\begin{figure}[t]
        \centering
        \begin{quantikz}
\lstick{$\ket{\psi}$} & & &  \gate[wires=2]{e^{-i\frac{t}{2}P\otimes Z}} &  & &  & \\
\lstick{$\ket{0}_a$} & &\targ{} &  & \targ{} & &  & \\
\lstick{$\ket{0}_c$} & \gate{R^\dagger_x(\pi/2)} & \ctrl{-1} &  & \ctrl{-1} & \gate{R_x(\pi/2)} & \gate{H} & \rstick{$\ket{0}_c$} 
\end{quantikz}
        \caption{Circuit that implements the non-unitary prescription in Eq.~\eqref{eq:exp_Pauli_non_unitary}, by exploiting the initialization and measurement of the ancilla qubit $c$.}
        \label{fig:exp_P_non_unitary}
\end{figure}
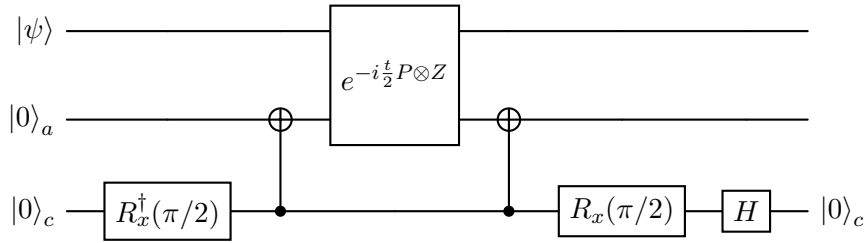

Our approach naturally allows for the construction of non-unitary trigonometric gates, in which the factor of $i$ is removed from the exponent. These operators will be used in the state-preparation algorithm discussed in section~\ref{sec:QITE}. This can be achieved by converting Eq.~\eqref{eq:exp_Pauli_unitary} to a non-unitary expression using an additional ancillary control qubit $c$. Specifically, the qubit is coupled via a controlled-$Y$ operation, followed by the application of a Hadamard gate and post-selection on the appropriate ancilla outcome. 
\begin{equation}
    \bra{0}_c\,H_ce^{-i\frac{t}{2}P\otimes Z_a\otimes Y_c}\ket{\psi}\ket{0}_a\ket{0}_c\biggr\rvert_{\tan(t/2)\mapsto \tanh(t/2)} = \frac{1}{\cosh(t/2)\sqrt{2(1+\tanh^2(t/2))}}e^{-\frac{t}{2}P}\ket{\psi}\ket{0}_a.
    \label{eq:exp_Pauli_non_unitary}
\end{equation}
The circuit is shown in Fig.~\ref{fig:exp_P_non_unitary}.
Note that this step keeps the qumode sector of the circuit unaffected, as it is purely an exact transformation of the single qubit operations. The replacement $\tan(t/2)\mapsto \tanh(t/2)$ is used to conveniently redefine the parameter $t$, which is transformed non-trivially by the non-unitary operation. The above allows us to create the non-unitary trigonometric operators
\begin{equation}
    e^{- t \cos\hat{A}}\quad\text{and}\quad e^{- t \sin\hat{A}},
\end{equation}
up to higher order corrections in $t$, which can be systematically improved using high-order Trotter decompositions as for the corresponding unitary gates discussed before.

\section{Quantum simulation of the sine-Gordon model}
\label{sec:sineGordon}

After constructing the trigonometric continuous-variable gates introduced above, we now turn to an application in the quantum simulation of interacting field theories. As a representative example, we focus on the sine-Gordon model, which captures a wide range of physical phenomena in high-energy, nuclear, and condensed matter physics, as discussed in the introduction. We begin by reviewing the lattice-discretized Hamiltonian formulation of the model and describe its implementation using continuous–discrete hybrid quantum gates. We then present classical simulations of small lattice systems, investigating several quantities, including real-time evolution, ground-state preparation, time-dependent two-point correlation functions, and quantum kink profiles.

\subsection{Discretized Hamiltonian}

An exemplary quantum field theory whose Hamiltonian requires the trigonometric continuous-variable gates introduced above is the well-known sine-Gordon model in 1+1 dimensions. The Lagrangian density in the continuum is given by~\cite{Zamolodchikov:1995xk}
\begin{equation}
\mathcal{L}_{\text{sG}}
=
\frac{1}{2}\,\partial_\mu \phi\,\partial^\mu \phi
-
\frac{m^2}{\beta^2}\left(1-\cos(\beta \phi)\right),
\end{equation}
where, $m$ and $\beta$ are the parameters of the theory. 

The spectrum of the quantum sine-Gordon Hamiltonian strongly depends on the value of the coupling parameter $\beta$. 
In the region $0 < \beta^2 < 4\pi$, the spectrum contains solitons (kinks), antisolitons (antikinks), and kink–antikink bound states known as breathers \cite{ Zamolodchikov:1978xm}. 
For $4\pi < \beta^2 < 8\pi$, the breathers disappear and the spectrum consists solely of kinks and antikinks. 
The special point $\beta^2 = 4\pi$ corresponds to a free massive fermion theory \cite{Coleman:1974bu}. 
At the Coleman point, $\beta^2 = 8\pi$, the cosine potential becomes marginal in the renormalization group sense, and for $\beta^2 > 8\pi$ it is irrelevant, such that the theory flows to a gapless free boson in the continuum limit \cite{Coleman:1974bu, Zamolodchikov:1978xm}. 
These two points, $\beta^2 = 4\pi$ and $\beta^2 = 8\pi$, are particularly important as they mark qualitative changes in the spectrum and the low-energy behavior of the model.

The corresponding Hamiltonian, discretized on a spatial lattice~\cite{Kogut:1974ag}, can be written as
\begin{equation}
    H_{\text {sG }}=\sum_{n=0}^{L-1} a\left[\frac{1}{2} \pi_n^2+\frac{1}{2}\left(\frac{\phi_{n+1}-\phi_n}{a}\right)^2+\frac{m^2}{\beta^2}\left(1-\cos \left(\beta \phi_n\right)\right)\right]\,,
\end{equation}
where $a$ is the lattice spacing, which we set to unity from here on. Throughout this work, we use periodic boundary conditions. The generalized momentum $\pi_n$  at lattice site $n$ and the scalar field $\phi_n$ satisfy the bosonic algebra
\begin{equation}\label{eq:commutator}
    [\phi_n,\pi_{n'}] = i\delta_{nn'}\,.
\end{equation}
Except for the potential term, this Hamiltonian is the same as for scalar $\phi^4$ field theory, which was discussed in the context of continuous-variable quantum computing in Refs.~\cite{Marshall:2015mna,Briceno:2023xcm,Abel:2024kuv,Abel:2025pxa,Abel:2025zxb,Gupta:2025xti}. The commutation relation in Eq.~(\ref{eq:commutator}) suggests a natural mapping of the field and momentum variables to the quadrature operators of a single qumode per site, i.e. $\phi_n\mapsto \hat{x}_n,\; \pi_n\mapsto\hat{p}_n$. However, since we are ultimately interested in the unitary time evolution operator, we can choose to first diagonalize the quadratic part of the Hamiltonian so that we only have local (single-mode) gates. A particularly convenient way of simulating the Trotterized time evolution is via the Bloch-Messiah decomposition~\cite{Bloch:1962zj,Braunstein:1999di}. While this procedure has been described in the literature, we review it here to establish the notation that is needed to determine the relevant trigonometric gates to simulate this model. We start by considering the gradient term. Writing the sum symmetrically, we find
\begin{align}
    H_{\rm grad}&\,=\frac{1}{2}\sum_{n=0}^{L-1}\left(\frac{\phi_{n+1}-\phi_n}{a}\right)^2=\frac{1}{2} \sum_{n, m}^{L-1} \phi_n K_{n m} \phi_m\,,
\end{align}
where 
\begin{equation}
    K_{n m}=2 \delta_{n m}-\delta_{n, m+1}-\delta_{n, m-1} -\delta_{0,L-1}-\delta_{L-1,0}\,.
\end{equation}
Here, $K$ is a circulant tridiagonal Toeplitz matrix whose eigenvectors are given by
\begin{equation}
    \boldsymbol{v}^{(s)} = \frac{1}{\sqrt{L}}\begin{pmatrix}
        1,w^s,w^{2s},\dots,w^{(L-1)s}
    \end{pmatrix},\;\; \text{with}\;\; w = e^{-i\frac{2\pi}{L}}\,,
\end{equation}
and eigenvalues
\begin{equation}
    \omega^2_s  = 2-2\cos\left(\frac{2\pi}{L}s\right),\quad s = 0,\dots,L-1\,.
\end{equation}
This is exactly the discrete Fourier transform (DFT) of the fields
\begin{equation}
\begin{split}
    \tilde{\phi}_s \equiv \boldsymbol{v}^{(s)}\cdot\boldsymbol{\phi} = \sum_{n=0}^{L-1}v^{(s)}_n\phi_n,\quad \tilde{\pi}_s\equiv \boldsymbol{v}^{(-s)}\cdot\boldsymbol{\pi} = \sum_{n=0}^{L-1}v^{(s)*}_n\pi_n.
\end{split}
\end{equation}
However, this basis is not convenient as the eigenvectors are not Hermitian. The DFT can be made real by noticing 
that $\tilde{\phi}_{L-s} = \tilde{\phi}_s^\dagger$. Therefore, for $s\neq 0$ it is more convenient to define
\begin{equation}
    \begin{split}
        \tilde{\Phi}_{s}&\equiv \frac{\tilde{\phi}_s + \tilde{\phi}_{L-s}}{\sqrt{2}} = \sqrt{2}\,\text{Re}[\tilde{\phi}_s],\quad
        \tilde{\Phi}_{L-s}\equiv \frac{\tilde{\phi}_s - \tilde{\phi}_{L-s}}{i\sqrt{2}} = \sqrt{2}\,\text{Im}[\tilde{\phi}_s]\,,
    \end{split}
\end{equation}
and similarly
\begin{equation}
    \begin{split}
        \tilde{\Pi}_{s}&\equiv \frac{\tilde{\pi}_s + \tilde{\pi}_{L-s}}{\sqrt{2}}= \sqrt{2}\text{Re}[\tilde{\pi}_s],\quad 
        \tilde{\Pi}_{L-s}\equiv -\frac{\tilde{\pi}_s - \tilde{\pi}_{L-s}}{i\sqrt{2}} = -\sqrt{2}\text{Im}[\tilde{\pi}_s]\,.
    \end{split}
\end{equation}
The component with $s = 0$ is already Hermitian
    \begin{equation}
        \tilde{\Phi}_0 = \tilde{\phi}_0 = \tilde{\phi}_0^\dagger,\quad \tilde{\Pi}_0 = \tilde{\pi}_0 = \tilde{\pi}_0^\dagger.
    \end{equation}
For even $L$, there is another special case $L-s=s$, which means $s = L/2$. This is referred to as the Nyquist mode. In this case, we have
    \begin{equation}
        \tilde{\Phi}_{\frac{L}{2}} = \tilde{\phi}_{\frac{L}{2}} = \tilde{\phi}_{\frac{L}{2}}^\dagger,\quad \tilde{\Pi}_{\frac{L}{2}} = \tilde{\pi}_{\frac{L}{2}} = \tilde{\pi}_{\frac{L}{2}}^\dagger.
    \end{equation}
To summarize, we work with the following transformed operators
        \begin{equation}
        \begin{split}
            \tilde{\Phi}_{s} & = \begin{cases}
                \frac{1}{\sqrt{L}}\sum_{n=0}^{L-1}\phi_n,\\
                 \sqrt{\frac{2}{L}}\sum_{n=0}^{L-1}\cos\left(\frac{2\pi}{L}ns\right)\phi_n, \\
                 \frac{1}{\sqrt{L}}\sum_{n=0}^{L-1}(-1)^{n}\phi_n,\\
                 \sqrt{\frac{2}{L}}\sum_{n=0}^{L-1}\sin\left(\frac{2\pi}{L}ns\right)\phi_n
            \end{cases},\; \tilde{\Pi}_{s}  = \begin{cases}
                 \frac{1}{\sqrt{L}}\sum_{n=0}^{L-1}\pi_n,\quad &(s=0),\\
                \sqrt{\frac{2}{L}}\sum_{n=0}^{L-1}\cos\left(\frac{2\pi}{L}ns\right)\pi_n, & (0<s<L/2),\\
                \frac{1}{\sqrt{L}}\sum_{n=0}^{L-1}(-1)^{n}\pi_n,\quad &(s=L/2),\\
                \sqrt{\frac{2}{L}}\sum_{n=0}^{L-1}\sin\left(\frac{2\pi}{L}ns\right)\pi_n, & (L/2<s<L).
            \end{cases}
        \end{split}
    \end{equation}
These are  Hermitian operators satisfying the canonical bosonic algebra $[\tilde{\Phi}_{s},\tilde{\Pi}_{r}] = i\delta_{sr}$.
Therefore, the entire quadratic part of the Hamiltonian can be written as
\begin{equation}
    H_{\rm quad}= \begin{cases}
        \frac{1}{2}\tilde{\Pi}_0^2+\frac{1}{4}\sum_{s=1}^{L-1}\tilde{\Pi}_s^2 + \frac{1}{4}\sum_{s=1}^{L-1}\omega_s^2\tilde{\Phi}_s^2,\quad &\text{odd }$L$,\\
        \frac{1}{2}\tilde{\Pi}_0^2+\frac{1}{2}\tilde{\Pi}_{\frac{L}{2}}^2 + \frac{1}{2}\omega^2_{\frac{L}{2}}\tilde{\Phi}_{\frac{L}{2}}^2+\frac{1}{4}\sum_{s=1,s\neq\frac{L}{2}}^{L-1}\tilde{\Pi}_s^2 + \frac{1}{4}\sum_{s=1,s\neq\frac{L}{2}}^{L-1}\omega_s^2\tilde{\Phi}_s^2,\quad &\text{even }$L$
    \end{cases},
    \label{eq:quad_hamiltonian}
\end{equation}
with $\omega_s^2 = \omega_{L-s}^2$ when $\ s\neq 0$. In this basis, the potential term gets modified as
\begin{equation}
    H_{\rm pot} = \frac{m^2}{\beta^2}\sum_{n=0}^{L-1}\left(1-\cos\left(\beta \sum_{s=0}^{L-1}V_{ns}\tilde{\Phi}_s\right)\right) = \frac{m^2}{\beta^2}\sum_{n=0}^{L-1}\left(1-\cos\left(\beta \bm{V}_n\cdot\tilde{\bm{\Phi}}\right)\right),
\end{equation}
where $V$ is the matrix that changes the operators in the real DFT basis back to the physical space
\begin{equation}
    V_{ns} = \begin{cases}
                \frac{1}{\sqrt{L}},\quad &(n=0),\\
                \sqrt{\frac{2}{L}}\cos\left(\frac{2\pi}{L}ns\right), & (0<n<L/2),\\
                \frac{(-1)^s}{\sqrt{L}},\quad &(n=L/2),\\
                \sqrt{\frac{2}{L}}\sin\left(\frac{2\pi}{L}ns\right), & (L/2<n<L).
            \end{cases}
            \label{eq:DFT_matrix}
\end{equation}

\subsection{Time evolution}

We implement the unitary time evolution operator in two steps: we first consider the quadratic and potential contributions separately, and then we adopt the (first-order) Trotter-Suzuki formula to obtain the desired time $t$,
\begin{equation}
    U(t)\equiv e^{-i t \left( H_{\rm quad} + H_{\rm pot}\right)} = \lim_{N\to\infty}\prod _{n=1}^Ne^{-i \frac{t}{N}  H_{\rm quad} }e^{-i \frac{t}{N}  H_{\rm pot}}.
\end{equation}
For the quantum simulation, it is natural to represent the bosonic degrees of freedom as
\begin{equation}
    \tilde{\Pi}_s\to \hat{p}_s,\quad \tilde{\Phi}_s\mapsto\hat{x}_s.
\end{equation}
Doing so reduces the quadratic Hamiltonian to a sum of local terms of the form 
\begin{equation}
    H_{\rm quad} = \sum_{s=0}^{L-1}\left(A_s\hat{p}_s^2 + B_s\hat{x}_s^2\right)\equiv \sum_{s=0}^{L-1}H_{\rm quad,s}\,,
    \label{eq:quadraticHam_sumlocal}
\end{equation}
where the coefficients can be found by comparing to Eq.~\eqref{eq:quad_hamiltonian}.
The evolution operator of a local quadratic Hamiltonian of the form $H_{\rm quad,s}$ in Eq.~\eqref{eq:quadraticHam_sumlocal} can be realized in terms of two squeezing and one rotation gate 
\begin{equation}
    e^{-it H_{\rm quad,s}} = S_s(r_s)R_s(\Omega_s t)S_s^\dagger(r_s)\,,
\end{equation}
with
\begin{equation}
    r_s = \frac{1}{4}\ln\left(\frac{A_s}{B_s}\right),\quad \Omega_s = 2\sqrt{A_sB_s}\,.
\end{equation}
The exact full unitary evolution operator of the quadratic part is
\begin{equation}\label{eq:SRS1}
    U_{\rm quad}(t) = e^{-i \frac{t}{2}\hat{p}_0^2}\bigotimes_{s=1}^{L-1}S_s(r_s)R_s(\Omega_st)S_s^\dagger(r_s).
\end{equation}
Notice that we have excluded the zero mode with vanishing normal frequency ($\omega^2_0 = 0$) from the $SRS^\dagger$ decomposition, as it would be equivalent to an infinite squeezing. Instead, it can be implemented via a quadratic phase gate, which leads to the prefactor in Eq.~(\ref{eq:SRS1}).

\begin{figure}[t]
    \centering
    \includegraphics[width=0.7\linewidth]{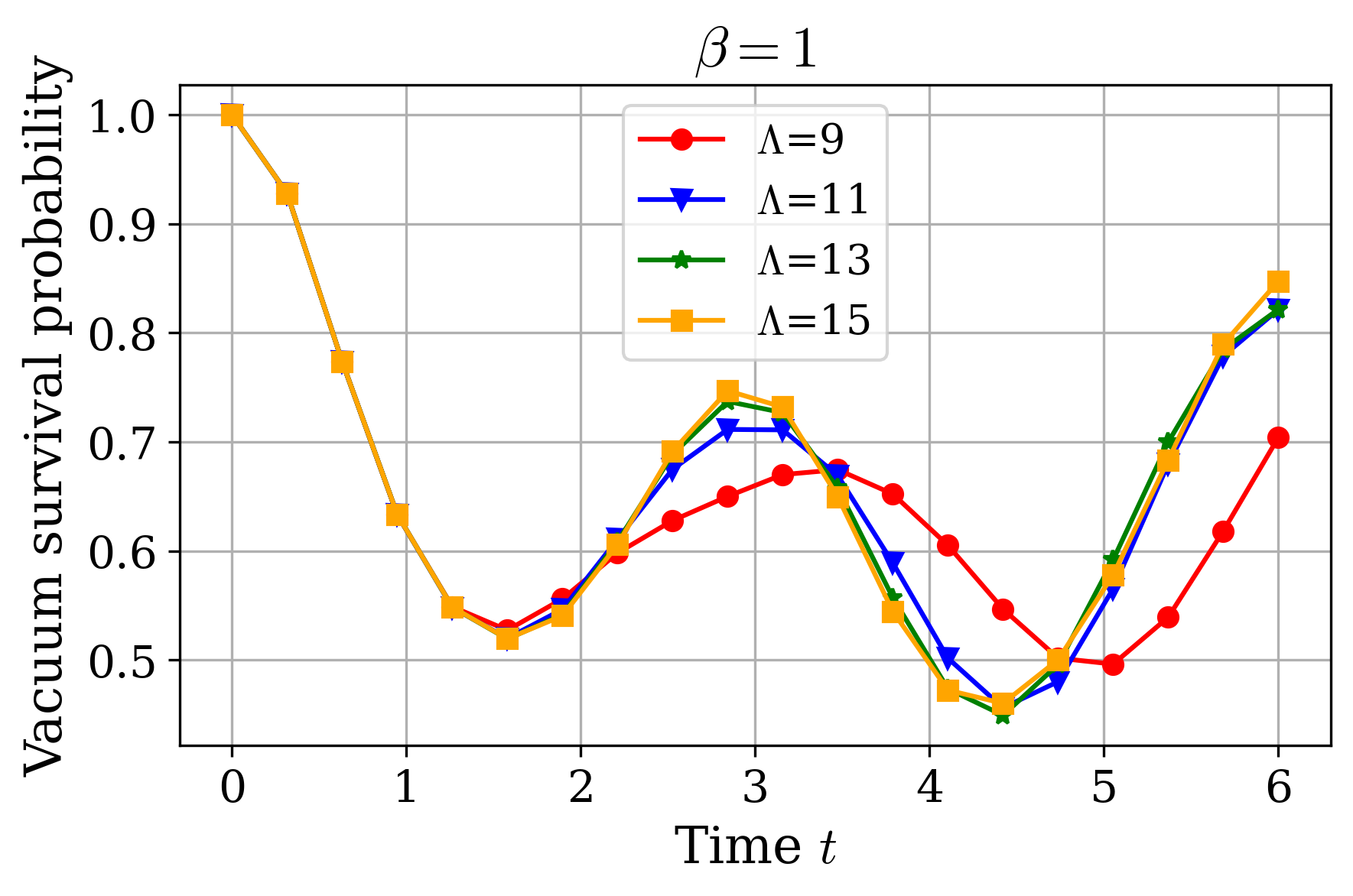}
    \caption{Survival probability of the free vacuum $\ket{n=0}^{\otimes L}$ for $L=3$ lattice sites as a function of time for representative model parameters. For the classical simulations, we choose different cutoffs for the local Hilbert space as indicated in the figure. The markers indicate Trotter steps.}
    \label{fig:Vacuum_survival_L3_beta1_plot}
\end{figure}
Since we work in Fourier space, the time evolution operator corresponding to the potential term in the Hamiltonian is expressed exactly as the product of $L$ cosine gates, each having a sum of $L$ quadratures as their argument. Specifically, it is
\begin{equation}
    U_{\rm pot}(t) = \bigotimes_{n=0}^{L-1}e^{it\frac{m^2}{\beta^2}\cos\left(\beta \bm{V}_n\cdot\,\hat{\bm{x}}\right)}
\end{equation}
The ancillary qubit resources to build the cosine of a single quadrature or of a sum of many are always the same with our procedure. Adding commuting terms as arguments of the cosine amounts to simply adding more control displacements on the ancilla qubit in parallel.

Alternatively, we could also express the unitary evolution operator in terms of the physical basis by undoing the change of basis $V$. In that case, it is convenient to keep the $SRS^\dagger$ decomposition of the quadratic contribution and transform it as
\begin{equation}
    U_{\rm quad}(t)\mapsto V^TU_{\rm quad}(t)V\,,
\end{equation}
where $V$ can be implemented with single-phase rotations and beam-splitters \cite{Girouard:2025enk, Marshall:2015mna}. In the physical basis, the potential term is simply the product of $L$ cosine gates, each one having only the corresponding site quadrature $\hat{x}_n$ as argument. For the numerical results discussed below, we verified that both approaches give identical results up to finite truncation effects of the qumode Hilbert space.

For the following analysis, we only consider the theory in Fourier space. In Fig.~\ref{fig:Vacuum_survival_L3_beta1_plot}, we show the survival probability of the free vacuum $\ket{n=0}^{\otimes L}$ for three lattice sites ($L=3$), setting $m = 1$ and $\beta = 1$ for four different truncations of the local qumode Hilbert space, which is necessary for the classical simulation. We observe good convergence of the survival probability as the local cutoff $\Lambda$ is increased. In particular, the results for $\Lambda = 11, 13,$ and $15$ show only small quantitative differences over the time window considered, indicating that $\Lambda = 11$ already provides a satisfactory approximation for the range of time evolution studied here. All simulations are performed using sufficiently small Trotter steps such that discretization errors are negligible compared to truncation effects.

\begin{figure}[t]
    \centering
    \includegraphics[width=0.48\linewidth]{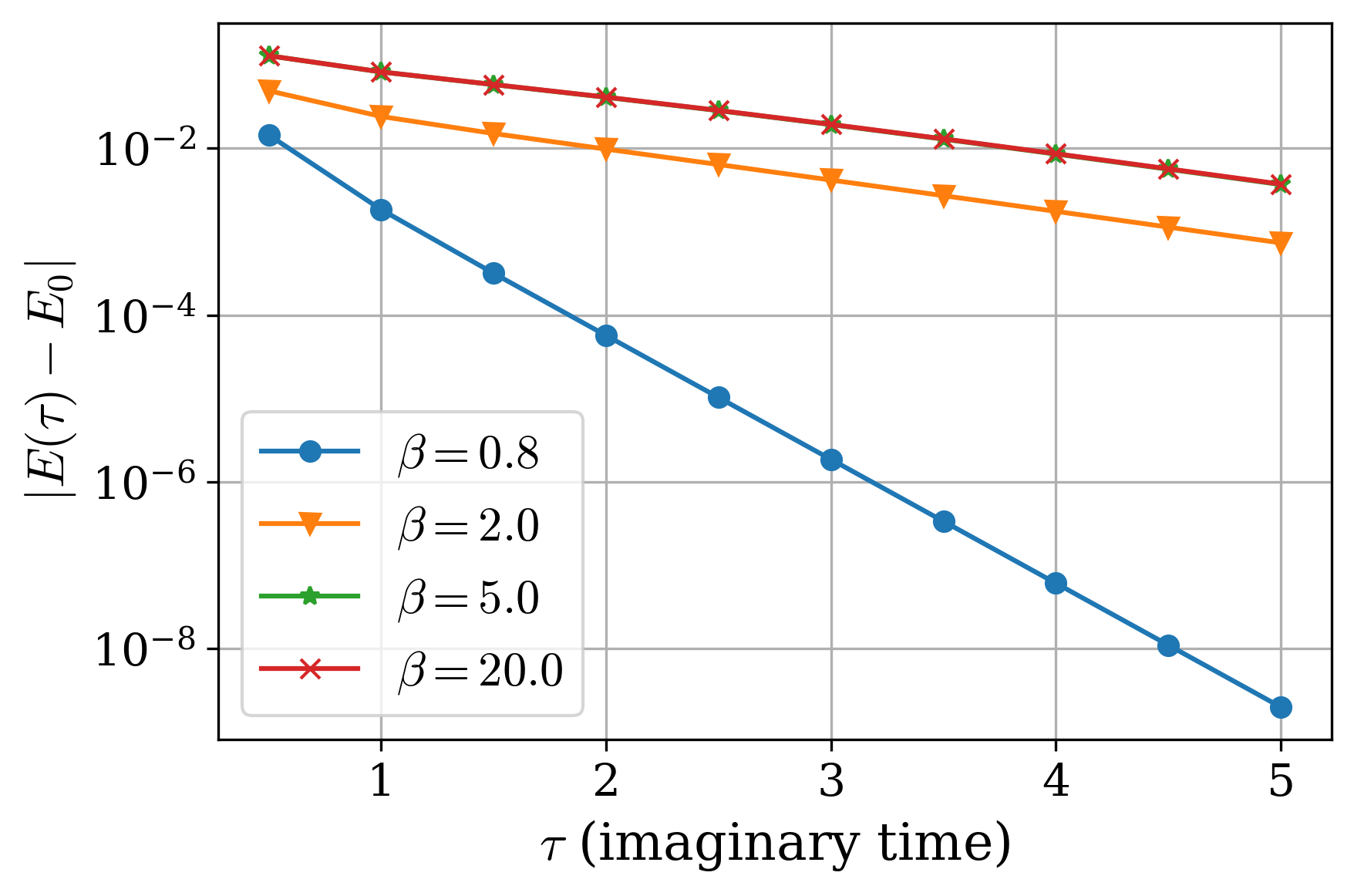}
    \includegraphics[width=0.48\linewidth]{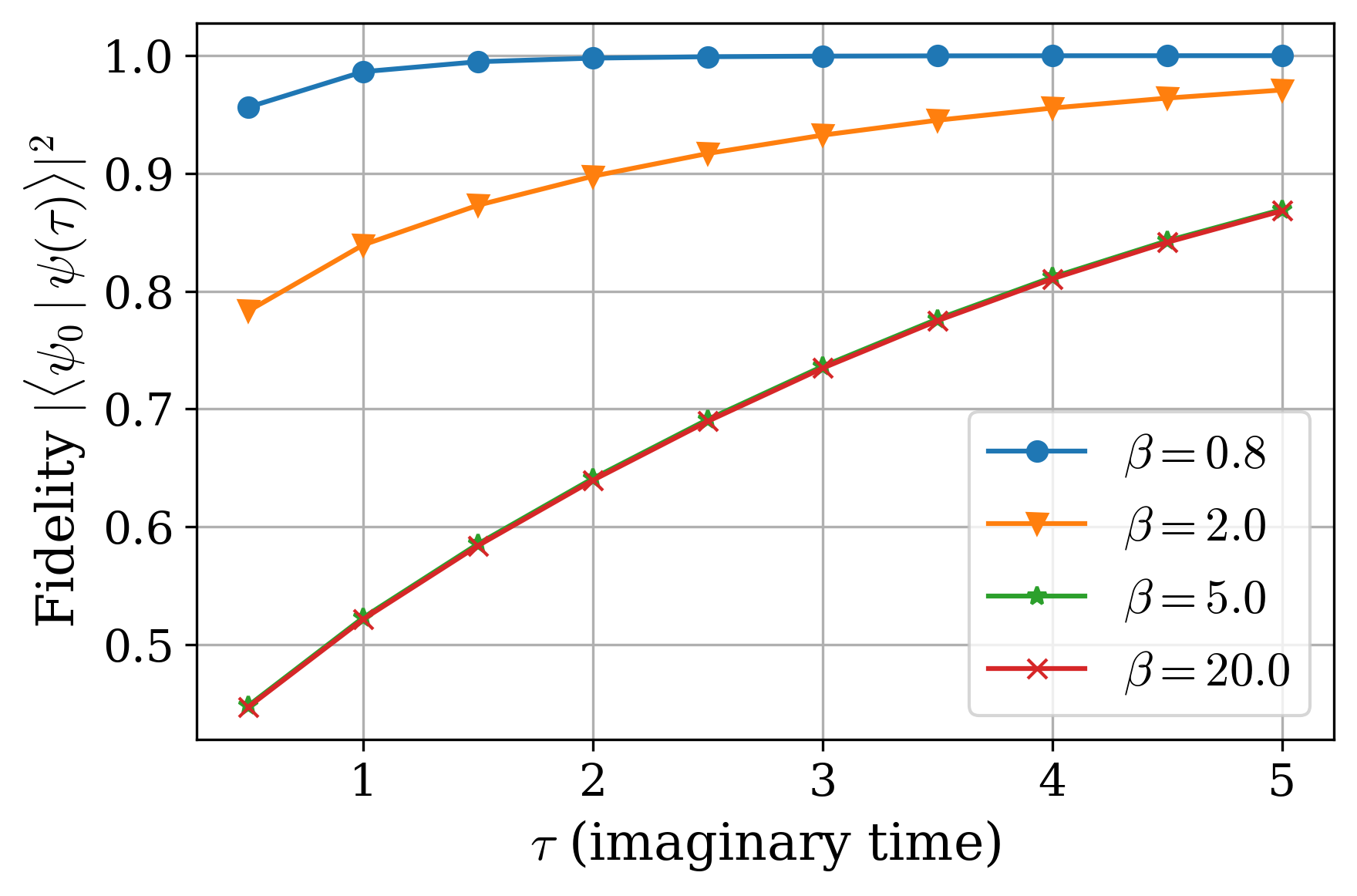}
    \caption{Comparison between the true ground state and the result obtained using the QITE algorithm. We quantify the agreement in terms of the ground state energy (left) and the fidelity (right). The markers indicate imaginary time steps associated with the QITE algorithm.}
    \label{fig:QITE_convergence}
\end{figure}

\subsection{State preparation with QITE}
\label{sec:QITE}

The ground state of any gapped theory can be found by exploiting the rapid decay of the non-unitary imaginary-time evolution~\cite{Motta2020,PhysRevA.105.012412, Ale:2025sxz},
\begin{equation}\label{eq:general_Qite}
    \frac{\bra{\phi}e^{-2\tau H} H\ket{\phi}}{\bra{\phi}e^{-2\tau H}\ket{\phi}}= E_0 + \mathcal{O}(e^{-2\tau (E_1-E_0)}),
\end{equation}
where $\ket{\phi}$ is an arbitrary state with non-vanishing overlap with the ground state.

We employ the circuits described in section~\ref{sec:nonU_trigs} to implement the non-unitary time evolution. As before, we consider a lattice with $L=3$ sites and impose a local cutoff of $\Lambda=11$. In the left panel of Fig.~\ref{fig:QITE_convergence}, we show the rapid convergence of the QITE algorithm toward the ground-state energy obtained from exact diagonalization of the Hamiltonian. We fix $m=1$ and consider several values of $\beta \in \lbrace 0.8,2,5,20\rbrace $. The imaginary-time step is chosen as $\Delta\tau=0.5$, and we employ $10$ Trotter steps. We observe that satisfactory accuracy is already achieved with fewer steps, particularly for smaller values of $\beta$. In contrast, larger values of $\beta$ require longer imaginary times to converge to the correct ground-state energy with comparable accuracy. The fidelity with respect to the exact ground state is shown in the right panel of Fig.~\ref{fig:QITE_convergence}, where a similar trend is observed. The convergence to the ground state is generally slower for larger values of $\beta$, which is consistent with a reduced spectral gap of the Hamiltonian. 

\subsection{Time-dependent two-point correlation function}

Next, we simulate time-dependent correlation functions of the sine-Gordon model. We denote the ground state of the lattice sine-Gordon model by $\ket{\Omega} = \ket{\Omega(m,\beta)}$. Given a local operator ${\cal O}(x,t)$, we define the corresponding time-dependent two-point connected ground state expectation value
\begin{equation}
    G^{(O)}_c(x,y,t) = 
    \bra{\Omega} {\cal O}(x,t){\cal O}(y,0) \ket{\Omega} 
    - \bra{\Omega} {\cal O}(x,0)\ket{\Omega} \bra{\Omega} {\cal O}(y,0) \ket{\Omega},
\end{equation}
with
\begin{equation}
    {\cal O}(x,t) = e^{i Ht}{\cal O}(x,0)e^{-iHt}.
\end{equation}
A particularly interesting observable is the so-called vertex operator ${\cal O}(x,t) = \exp\lbrace i \alpha \phi(x,t)\rbrace$, where $\alpha\in\mathbb{R}$ is a constant, from which we can compute the two-point function 
\begin{equation}
    G_c(n,k,t) = 
    \bra{\Omega} e^{i \alpha \phi_n(t)} e^{-i \alpha \phi_k(0)} \ket{\Omega} 
    - \bra{\Omega} e^{i \alpha \phi_n(0)} \ket{\Omega} \bra{\Omega} e^{-i \alpha \phi_k(0)} \ket{\Omega}.
    \label{eq:vertex_corr_GS}
\end{equation}
For the three-site lattice ($L = 3$), we consider $n = 0$ and $k = L - 1$.
This operator measures the connected correlations of vertex operators at different lattice sites and times. Vertex operators $e^{i \alpha \phi_n}$ are of central importance in the sine-Gordon model because they create and annihilate soliton-like excitations and encode the non-perturbative structure of the theory, including the topological sectors and form factors. The connected correlator isolates the genuine two-point correlations by subtracting the product of the one-point functions, which would otherwise include trivial disconnected contributions.

Analytical results for such correlators are known in the continuum sine-Gordon model in certain regimes, particularly using form-factor expansions and the integrable structure of the model~\cite{Lukyanov:1996jj}. For general lattice parameters and finite systems, exact expressions are not available, and one typically relies on numerical approaches. 

\begin{figure}[t]
    \centering
    \includegraphics[width=0.7\linewidth]{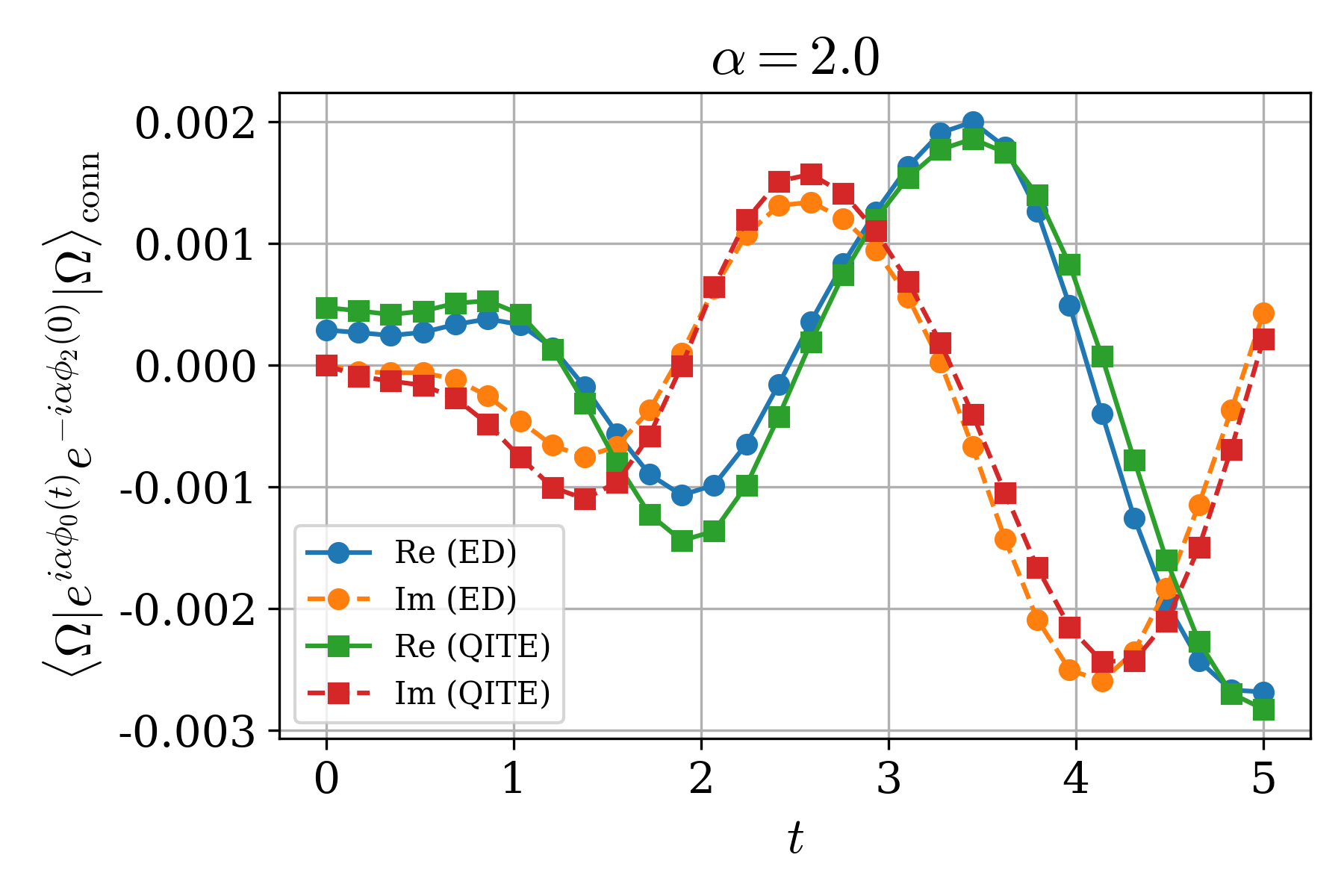}
    \caption{Connected vertex correlator $G_c(0,L-1,t)$ for $m=1$, $\beta = \alpha = 2$, and local Fock cutoff $\Lambda=11$ between site $0$ and $L-1$ with $L=3$. The result using exact diagonalization (ED) is compared with the approximated ground state (with fidelity 0.971) using QITE. The agreement can be improved with a higher fidelity QITE prepared ground state.}
    \label{fig:ConnectedVertex}
\end{figure}

In this work, we approximate the true ground state by the one obtained with QITE and use it to compute the connected two-point function in Eq.~\eqref{eq:vertex_corr_GS}. In Fig.~\ref{fig:ConnectedVertex}, we show a comparison between the correlator at sites $0$ and $L-1$, with $L=3$, evaluated on the exact ground state (via exact diagonalization) and the one obtained with QITE using 10 Trotter steps and $\Delta\tau = 0.5$. Both calculations are performed for $m = 1, \beta = \alpha = 2$ and a local Fock cutoff of $\Lambda=11$, corresponding to a ground state fidelity of approximately $0.971$.

The results demonstrate that a QITE-prepared approximate ground state is already sufficient to reproduce the main features of the connected two-point function of vertex operators, showing good agreement with exact diagonalization even for a modest circuit depth and a finite imaginary-time step. The remaining discrepancies are consistent with the finite Trotter step and the imperfect ground-state fidelity. The ability to capture nontrivial, nonlocal correlations using a relatively shallow imaginary-time evolution highlights the potential of QITE as a viable tool for studying interacting quantum field theories on NISQ-era hardware. The present results are restricted to small system sizes and relatively large time steps. However, systematic improvements in hardware capabilities and algorithmic refinements are expected to enable simulations of larger lattices and finer imaginary-time discretizations, potentially extending these studies into regimes that are increasingly challenging for classical methods.

\subsection{Quantum kink profile}

In the sine-Gordon model, a kink is a configuration in which the field interpolates between two adjacent minima of the cosine potential \cite{Manton:2004tk}. See Ref.~\cite{Milsted:2020jmf}, where kink anti-kink scattering was studied for an extension of the trasnverse Ising spin chain \cite{PhysRevLett.120.206403} using tensor networks. 

\begin{figure}[t]
    \centering
    \includegraphics[width=0.7\linewidth]{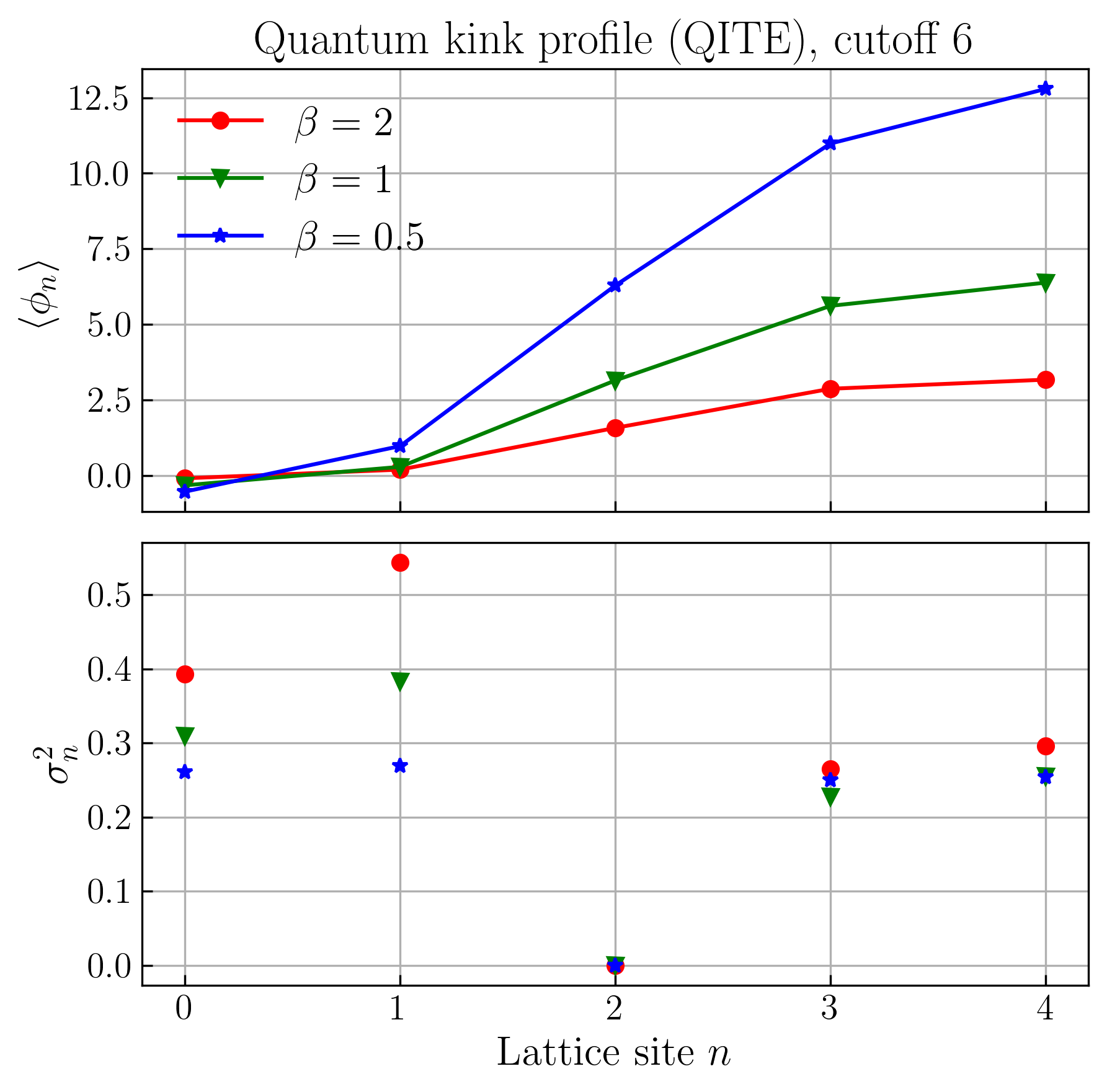}
    \caption{Expectation value (top) and variance (bottom) of the quantum kink profile using QITE for the ground state.}
    \label{fig:KinkProfile_QITE}
\end{figure}

On a finite lattice, the kink configuration can be enforced by imposing boundary conditions $\phi_0 = \phi_{\rm left}$ and $\phi_{L-1} = \phi_{\rm right}$, which we separate by $2\pi/\beta$ to obtain a kink with topological charge $+1$. In the real Fourier basis, the lattice field operators are expanded as $\phi_n = \sum_{s=0}^{L-1} V_{ns} \tilde{\Phi}_s$, where $V_{ns}$ is the real DFT matrix in Eq.~\eqref{eq:DFT_matrix}. The kink is imposed by first computing the classical lattice kink $\{\phi_n^{\rm cl}\}$, i.e., the static field configuration that minimizes the classical lattice static energy
\begin{equation}
    E_{\rm cl}[\phi] = \sum_{n=0}^{L-1} \frac{1}{2} (\phi_{n+1} - \phi_n)^2 + \frac{m^2}{\beta^2}\sum_{n=0}^{L-1} \big(1 - \cos(\beta \phi_n)\big),
\end{equation}
subject to the boundary conditions $\phi_0 = \phi_{\rm left}$ and $\phi_{L-1} = \phi_{\rm right}$. The Fourier-space shift vector $\Phi_{\rm shift}$ is then obtained by projecting this classical solution onto the real Fourier basis via
\begin{equation}
    \Phi_{\rm shift,s} = \sum_{n=0}^{L-1} V_{sn} \, \phi_n^{\rm cl}, \qquad s = 0, \dots, L-1.
\end{equation}
The full quantum Hamiltonian is thus constructed using the shifted field operators $\phi_n^{\rm total} = \sum_s V_{ns} \tilde{\Phi}_s + \Phi_{\rm shift,n}$ in the cosine potential, while the quadratic part remains unchanged. Diagonalizing this Hamiltonian (or evolving it via QITE) yields a ground state whose expectation values $\langle \phi_n \rangle\equiv \bra{\Omega}\phi_n\ket{\Omega}$ form a quantum kink interpolating between the boundaries. The local quantum fluctuations are captured by the variances $\sigma_n^2 = \langle (\phi_n^{\rm dyn})^2 \rangle - \langle \phi_n^{\rm dyn} \rangle^2$.
In Fig.~\ref{fig:KinkProfile_QITE} we show the kink profile and variance for $L=5$, $m = 1$ and $\beta \in \lbrace 0.5, 1 , 2\rbrace$ for the local Fock cutoff $\Lambda = 6$. Given the relatively large number of lattice sites, we are limited to this small cutoff. The semi-classical limit is obtained as $\beta \to 0$, and, indeed, we find that the variances become smaller for decreasing values of $\beta$, confirming the sharper localization of the states. This behavior is consistent with the expected transition from a quantum to a classical regime, where fluctuations are suppressed and the system exhibits more well-defined, localized configurations.

Having explicit access to kink (and anti-kink) states opens the door to a broader range of investigations beyond static properties. In particular, it enables the computation of expectation values of local and nonlocal observables in topologically nontrivial sectors, allowing for a detailed characterization of kink-induced modifications to correlation functions and energy densities. These states provide a natural starting point for studying real-time dynamics, such as kink propagation, scattering, and responses to perturbations, which are essential for understanding nonequilibrium phenomena and the interplay between topology and dynamics in interacting quantum field theories.

\section{Conclusions and outlook}
\label{sec:conclusions}

In this work, we introduced trigonometric continuous-variable gates as a complementary universality paradigm for bosonic quantum computation within hybrid qubit–qumode architectures. By embedding non-polynomial bosonic operators into an enlarged Hilbert space using ancillary qubits, we developed deterministic and unitary circuits that implement trigonometric functions of arbitrary Hermitian qumode operators. This construction extends standard ancilla-based exponentiation techniques beyond Pauli strings and provides a systematic route to both unitary and non-unitary trigonometric gates, including those required for imaginary-time evolution.

We demonstrated the utility of this gate framework through a concrete quantum simulation of the lattice sine-Gordon model, an interacting field theory with a non-polynomial cosine potential. Using a hybrid qubit–qumode encoding, we showed how the discretized Hamiltonian can be efficiently mapped onto continuous-variable degrees of freedom, with trigonometric gates naturally capturing the structure of the interaction term. Within this framework, we prepared ground states via quantum imaginary-time evolution, simulated real-time dynamics, evaluated time-dependent vertex two-point correlation functions, and extracted quantum kink profiles under topological boundary conditions. These observables probe both equilibrium and dynamical properties of the model and illustrate how trigonometric gates provide direct access to non-perturbative features such as solitonic excitations and topological sectors.

Beyond the specific application studied here, our results highlight a broader conceptual point: trigonometric gates furnish a Fourier-based operator representation that complements the conventional polynomial (Taylor-based) approach to continuous-variable universality. For operators with periodic structure or global phase-space dependence, this representation can offer a more natural and potentially more resource-efficient description. Importantly, the two paradigms are not mutually exclusive; rather, they can be combined within hybrid circuits, with polynomial gates enabling the construction of trigonometric primitives and vice versa.

From a hardware perspective, the proposed gate constructions are compatible with leading hybrid quantum platforms. In trapped-ion systems, internal electronic states provide high-fidelity qubits while collective motional modes act as long-lived bosonic qumodes, with state-dependent forces enabling controlled displacements and multi-mode couplings. Superconducting circuit architectures similarly offer strong qubit–oscillator interactions through Josephson junctions coupled to microwave resonators, where conditional displacements and nonlinear ancilla-mediated operations are natively available. Cavity- and circuit-QED platforms further provide access to bosonic modes with engineered nonlinearities and long coherence times, making them particularly suitable for implementing the controlled operations required by the trigonometric gate protocols. While detailed resource estimates and noise analyses remain to be developed, the reliance on standard qubit–qumode primitives suggests that these gates can be integrated into existing experimental toolkits without requiring fundamentally new hardware capabilities.

Several directions for future work are immediate. On the algorithmic side, optimizing trigonometric gate decompositions, developing higher-order Trotter–Suzuki schemes, and exploring alternative compilation strategies tailored to specific hardware platforms will be essential for reducing circuit depth and accumulated errors. Extending the present framework to higher-dimensional field theories, where kinks generalize to line-like defects and domain walls, provides a natural path toward more complex and physically relevant simulations. More broadly, this work suggests that enriching the continuous-variable gate toolbox beyond polynomial operations by incorporating trigonometric primitives opens a qualitatively new avenue for quantum simulation, positioning hybrid qubit–qumode processors as a uniquely powerful platform for exploring non-perturbative quantum field theories and strongly interacting bosonic systems. In addition, we expect the trigonometric gates introduced here to find broader applications, including quantum simulations of condensed matter systems, quantum chemistry, and biological models.

Note added: Upon completion of this work, we became aware of the recent study in Ref.~\cite{chalermpusitarak2025programmablegenerationarbitrarycontinuousvariable}, where a closely related qubit-qumode approach was introduced to create trigonometric continuous-variable gates.

\section*{Acknowledgements}

We acknowledge support by DOE ASCR funding under the Quantum Computing Application Teams Program, NSF award DGE-2152168, and DOE Office of Nuclear Physics, Quantum Horizons Program, award DE-SC0023687. The research was supported by the DOE, Office of Science, Office of Nuclear Physics, Early Career Program under contract No. DE-SC0024358 and DE-SC0025881. 
The work of D.K. was supported by DOE, Office of Science, Office of Nuclear Physics under contract DE-FG88ER41450 and by the U.S. Department of Energy, Office of Science, National Quantum Information Science Research Centers, Co-design
Center for Quantum Advantage (C2QA) under Contract
No.DE-SC0012704. 
The authors would like to thank Stony Brook Research Computing and Cyberinfrastructure, and the Institute for Advanced Computational Science at Stony Brook University for access to the SeaWulf computing system, made possible by grants from the National Science Foundation ($\#$1531492 and Major Research Instrumentation award $\#$2215987), with matching funds from Empire State Development’s Division of Science, Technology and Innovation (NYSTAR) program (contract C210148). This work has been partially funded by the Eric \& Wendy Schmidt Fund for Strategic Innovation through the CERN Next Generation Triggers project under grant agreement number SIF-2023-004.

\bibliographystyle{utphys.bst}
\bibliography{bibliography}

\end{document}